# Emotional Speaker Identification using a Novel Capsule Nets Model

Ali Bou Nassif[1,a,*], Ismail Shahin[2,b], Ashraf Elnagar[3,c], Divya Velayudhan[1,d] Adi Alhudhaif[4,e], Kemal Polat[5,f]

[1]Department of Computer Engineering, University of Sharjah, Sharjah, UAE, 27272

[2]Department of Electrical Engineering, University of Sharjah, Sharjah, UAE, 27272

[3]Department of Computer Science, University of Sharjah, Sharjah, UAE, 27272

[4]Department of Computer Science, College of Computer Engineering and Sciences in Al-kharj, Prince Sattam bin Abdulaziz University, P.O. Box 151, Al-Kharj 11942, Saudi Arabia

[5]Bolu Abant Izzet Baysal University, Faculty of Engineering, Department of Electrical and Electronics Engineering, Bolu, Turkey

[a]anassif@sharjah.ac.ae,  [b]ismail@sharjah.ac.ae, [c]ashraf@sharjah.ac.ae, [d]divyavelayudhan@gmail.com, [e]a.alhudhaif@psau.edu.sa, [f]kpolat@ibu.edu.tr

*Corresponding Author: Ali Bou Nassif

***Abstract***—Speaker recognition systems are widely used in various applications to identify a person by their voice; however, the high degree of variability in speech signals makes this a challenging task. Dealing with emotional variations is very difficult because emotions alter the voice characteristics of a person; thus, the acoustic features differ from those used to train models in a neutral environment. Therefore, speaker recognition models trained on neutral speech fail to correctly identify speakers under emotional stress. Although considerable advancements in speaker identification have been made using convolutional neural networks (CNN), CNNs cannot exploit the spatial association between low-level features. Inspired by the recent introduction of capsule networks (CapsNets), which are based on deep learning to overcome the inadequacy of CNNs in preserving the pose relationship between low-level features with their pooling technique, this study investigates the performance of using CapsNets in identifying speakers from emotional speech recordings. A CapsNet-based speaker identification model is proposed and evaluated using three distinct speech databases, i.e., the Emirati Speech Database, SUSAS Dataset, and RAVDESS (open-access). The proposed model is also compared to baseline systems. Experimental results demonstrate that the novel proposed CapsNet model trains faster and provides better results over current state-of-the-art schemes. The effect of the routing algorithm on speaker identification performance was also studied by varying the number of iterations, both with and without a decoder network.

***Keywords***—*Capsule network;Convolutional Neural Network; Emotional speech; Speaker Identification*

# 1. INTRODUCTION

Speaker recognition has received increasing attention recently with increasing demands in various applications, e.g., security systems, biometric authentication, criminal investigation, and customer care (Alsharhan & Ramsay, 2019; B. Chen & Chen, 2013; Jahangir et al., 2021). Although we do not give significant consideration to the human capability of discriminating speakers using their voices alone, it has been proven to be an integral part of human–computer interactions and has been studied consistently (Shahin et al., 2021, 2022).

Speaker recognition methods can be text-dependent, where the speaker utters the same text for both training and testing, or text-independent, which does not involve restrictions on the text spoken during training and testing. Speaker recognition techniques can be broadly classified as speaker verification (SV) and speaker identification (SI) (Shahin et al., 2021). Similar to a password-entry system, speaker verification is used to determine a claimed speaker's identity (as true or false) using an audio sample (Shahin & Nassif, 2019; W. Wu et al., 2006). In contrast, given an audio sample, speaker identification attempts to identify the speaker from a pool of speakers. If the speaker whose identity is to be determined is from a set of known speakers, it is said to be a "closed-set" scenario; otherwise, it becomes an "open-set" scenario.

Several studies have proposed speaker identification techniques that have made significant contributions in the field (Farrell et al., 1994; Praveen Kumar et al., 2018); however, various conditions, e.g., noisy environments or the speaker's emotional, can affect the performance of such techniques (Bashirpour & Geravanchizadeh, 2018). The performance of these approaches is reduced significantly when the speech samples are obtained under suboptimal conditions, e.g., an emotional environment (J. H. L. Hansen & Patil, 2007), where the speech samples are emotional or highly expressive (Ghiurcau et al., 2011; Parthasarathy et al., 2017). Speech samples obtained under circumstances where the speaker is not influenced by any kind of emotion is referred to as 'neutral' speech samples, while those under any kind of emotion (such as happy, sad, anger and so on) are referred to as emotional speech. Emotional speech varies the acoustic features extracted from the samples.

Speaker identification models are often trained using neutral speech and perform well when tested on speech samples in a neutral environment, but not on speech samples recorded under emotional circumstances. However, human speech is more frequently emotional or expressive, which alters the speech characteristics of a speaker. These deviations in speech characteristics from neutral speech adversely affect model performance in the presence of such emotions (D. Li et al., 2005; Nassif et al., 2021). In summary, a neutral talking environment, where the speaker is not under stress or emotion yields better speaker identification performance compared to an emotional environment (e.g., when the speaker is angry, scared, happy, or afraid).

Nonetheless, recognizing the emotional aspect with linguistic statements is essential to establish a proper communication link in human–computer interfaces (Picard, 1999). However, this is a difficult problem when humans can also fail to accurately recognize emotion in a speech sample. Thus, speaker identification under stressful or emotional conditions is a challenging area of research.

CNN uses a pooling operation, which makes them insensitive to the spatial information of the input features, i.e., the position information of features in the time-frequency axis. The pooling function makes the CNN invariant and results in losing certain features. The capsule neural architecture was conceived to overcome these drawbacks (Hinton et al., 2011). Moreover, the capsule networks use extracted features such as MFCC to model the emotions. Furthermore, models are trained on Neutral only but tested in the emotional environment and this makes the accuracy higher in neutral.

Identifying an emotional speaker has numerous applications, e.g., identifying a person from emotional (anger or panic) speech samples during a crime investigation, identifying and assisting people in panic at emergency helpline centers and assessing the stress or frustration of callers at customer centers. Previous studies have focused on analyzing speaker identification under different emotional and stressful environments to realize optimal performance in robust human–computer interactions (Alluri et al., 2017; Fragopanagos & Taylor, 2005; Senthil Raja & Dandapat, 2010; S. Wu et al., 2011). Some studies (Scherer et al., 2000; T. Wu et al., 2005) have used emotional speech samples from speakers to train the model to improve speaker identification performance; however, this is not a feasible solution for practical applications.

Most state-of-the-art speaker identification models reported in the literature with identification performance greater than 90% utilized long utterances based on large speech corpora comprising speech segments that are nearly 2.5 minutes long (Gonzalez-Rodriguez, 2011). However, collecting long speech samples from speakers is unrealistic, but the results diminish considerably with shorter speech samples. Thus, speaker identification systems based on short utterances (Elnaggar & Arelhi, 2019; L. Li et al., 2016; Liu et al., 2018) have gained increasing attention in recent years. In addition, speaker identification research has primarily focused on English with little attention paid to Arabic (Saeed & Nammous, 2007).

Typical speaker identification systems involve feature extraction from speech samples followed by classification to identify the speaker. The increasing popularity of deep learning based techniques (J. et al., 2015; Nassif et al., 2019; Variani et al., 2014) in speech processing has motivated researchers to use convolutional neural networks (CNN) for SI (Lukic et al., 2016). However, CNN use a pooling operation, which makes them insensitive to the spatial information of the input features, i.e., the position information of features in the time-frequency axis. The pooling function makes the CNN invariant and results in losing certain features, which leads to the need for an extensive amount of training data to compensate the loss. However, acquiring sufficient amounts of training data can be difficult in practical application (Kwabena Patrick et al., 2019). The capsule neural architecture was conceived to overcome these drawbacks (Hinton et al., 2011), which was later reformed by introducing a dynamic routing algorithm (Sabour et al., 2017) to form capsule networks (CapsNets) (*End-to-End Speech Command Recognition with Capsule Network*, 2018).

The widespread applications of identifying speakers in stressed or emotional environments have encouraged researchers to investigate the cause from different perspectives, e.g., feature extraction and classification models. For example, Jawarkar et al. (Jawarkar et al., 2012) conducted an extensive study involving the performance of different features, i.e., MFCCs, line spectral frequencies, Teagerenergy-based mel-frequency cepstral coefficients, and temporal energy of sub-band cepstral coefficients and their combinations on a hybrid classifier based on a GMM. They also compared the results to those obtained by separate classifiers for SI for five different emotions involving 34 speakers in an Indian language. They concluded that fusing classifiers gives a better result for SI for both neutral and emotional speech data. Bao et al. (Bao et al., 2007) employed channel compensation methods to avoid the effects of emotions in the SI task after considering the resemblance it bears with the channel effect. They proposed the emotion attribute projection method, which reduced EER by 11.7%, and the linear fusion method based on GMM-UBM and SVM-EAP, which further reduced EER. Li et al. (D. Li et al., 2005) altered the features of speech samples in a neutral environment according to the prosodic parameters of speech samples in an emotional environment to enhance performance. In addition, Koolagudi et al. (Koolagudi et al., 2012) analyzed speaker identification in emotional environments using appropriate transformations to MFCCs and observed 16%–22% performance improvement when tested on a simulated database in an Indian language.

Ghiurcau et al. (Ghiurcau et al., 2011) studied the negative impact of the emotional state of a speaker on the performance of SI systems using the GMM. In a later study, they compared the results to those obtained using an SVM classifier on the Berlin speech database and concluded that models trained on neutral speech produced poor results on speech samples collected in an emotional environment. The results obtained varied with emotion (with anger and happiness scoring the least). The results were found to improve vastly when emotional speech was included in training, although this is not practical solution. Wu et al. (Zhaohui et al., 2006) explored ways to enhance speaker identification by studying techniques to alter prosodic features and utilizing the same to adapt the features from emotional speech. The authors used the EPST database, which comprising 14 different emotions by eight speakers. Chen and Yang (L. Chen & Yang, 2013) investigated modifying GMM weights to boost speaker identification and proposed the radial basis function neural network and exemplar-based sparse representation methods to synthesize emotional GMMs. In addition, Raja and Dandapat (Senthil Raja & Dandapat, 2010) assessed the variation in results produced by different features using vector quantization and the GMM as classifiers, and they recommended four compensation methods to alleviate the negative effects of emotional or expressive speech in SI. Parthasarathy and Busso (Parthasarathy & Busso, 2017) studied the effects on speaker identification in terms of arousal and valence, and they predicted dependable regions in emotional speech to improve speaker recognition performance.

Tirumala et al. (Tirumala et al., 2017) explored existing feature extraction techniques employed for SI and concluded that approaches based on Mel-frequency cepstral coefficients (MFCC) are the most frequently used. MFCCs have been used as feature inputs to the CNN architecture for SV in neutral environments (Salehghaffari, 2018). A deep learning architecture based on a 1D CNN with maxpooling and dense layers trained using MFCCs has been used for emotion recognition (De Pinto et al., 2020) on the RAVDESS dataset. In addition, Ali et al. (Ali et al., 2018) evaluated the performance of speaker recognition models on variable length audio when MFCCs are grouped with learned features using an Urdu dataset. Mackov et al. (Macková & Čižmár, 2014) used the I-vector approach to recognize speakers from emotional speech using a German emotional dataset (Emo-DB). MFCCs, log energy, and first- and second-order derivatives were used as features, and an overall accuracy of 80% was obtained. The influence of selecting different features for the i-vector approach has also been studied in the literature (Mackova et al., 2015). LPC, MFCC, and LPCC were used for front-end processing in extracting the i-vectors. The corresponding performance was evaluated, and overall accuracies of 80.33%, 72%, and 66.30% were observed for MFCC, LPC, and LPCC, respectively. Emotional speaker recognition in real-time situations was studied in the literature (Mansour et al., 2019) using different speech features (including MFCC, LPCC, RASTA-PLP, and MFCC-SDC) with i-vector modeling on IEMOCAP in both clean and noisy environments with different SNR levels. MFCC-SDC was found to yield the best results with an average accuracy of approximately 90% in a clean environment.

Meftah et al. (Meftah et al., 2020) exploited linearly spaced spectrograms to distinguish emotional speakers in Arabic and English using a fusion of a convolutional neural network and the long short-term memory architecture, and they based their study on the KSU Emotions, EPST, WEST POINT, and TIMIT databases. Recently, x-vectors were used (Pappagari et al., 2020) to analyze the effect of emotion in speaker verification in the IEMOCAP, MSP-Podcast, and Crema-D datasets. The results demonstrated that the performance of the systems degraded significantly when emotional circumstances during model training and testing were mismatched, they found that anger demonstrated the worst performance on all three datasets.

Despite the popularity of using CNNs to extract high level features, it has been found that the kernel-based convolution operation in a CNN can fail to capture certain features or its spatial information (Kwabena Patrick et al., 2019). In addition, the pooling operation renders them incapable of recognizing the spatial hierarchies between features. CapsNets, which are the latest addition to the family of deep learning models, have promised to overcome these drawbacks.

CapsNets have already been employed in speech processing (*End-to-End Speech Command Recognition with Capsule Network*, 2018) and emotion recognition (X. Wu et al., 2019) tasks. For example, a previous study (*End-to-End Speech Command Recognition with Capsule Network*, 2018) claimed that the position and relationship among features on the time-frequency axis, which differentiates the spectrograms corresponding to different speech samples, is overlooked by CNNs, and they observed that the performance of speech recognition systems on 1-s speech commands was improved using CapsNets compared to a baseline CNN. Sequential CapsNets have also been used (X. Wu et al., 2019) for speech emotion recognition to identify four emotions (neutral, happy, angry, and sad) over the benchmark corpus IEMOCAP, and the results were found to outperform CNN-LSTM models.

This goal of this study is to evaluate the performance of a modified CapsNet architecture to identify emotional speakers using text-independent short utterances and compare the results to baseline systems. This evaluation was performed by examining the results obtained on the Emirati Speech dataset, SUSAS English dataset, and RAVDESS (English) dataset. The models were trained using neutral speech samples, and then evaluated on emotional speech using utterances whose texts where not used for training.

This research aims to propose a novel CapsNets model for identifying speakers in an emotional environment. To the best of our knowledge, there has been no study to evaluate the performance of CapsNets for text-independent speaker identification using emotional speech. Our work evaluates the performance of our modified capsule net architecture on short utterances using three datasets – Arabic Emirati Speech Database, English Speech Under Simulated and Actual Stress (SUSAS) dataset and Ryerson Audio-Visual Database of Emotional Speech and Song (RAVDESS).

Our primary contributions are summarized as follows.

- This is the first work to employ CapsNets for text-independent speaker identification in an emotional environment using short utterances.

- The performance of the proposed CapsNet architecture is compared to a CNN in identifying speakers using emotional speech samples with MFCCs as the only input feature.

- Extensive experiments are performed to evaluate the performance of the proposed CapsNet architecture based on three distinct databases (English and Arabic) compared to baseline systems.

- Experimental analysis shows that the CapsNet model trains faster and provide better results over baseline systems.

The remainder of this paper is organized as follows. Section II provides a detailed explanation of CapsNets. Information about the datasets used in this study is given in Section III. The proposed CapsNet-based speaker identification model is discussed in Section IV, and Section V describes our experimental analysis and presents results. Finally, the paper is concluded in Section VI.

## 2. CAPSULE NETWORKS

CNNs are partially responsible for the increase of interest in deep learning models in recent years (Bunrit et al., 2019; Lukic et al., 2016). Convolutional layers, which form the first layers of a classical CNN, facilitate the extraction of features by applying variable-sized filters to the input to form feature maps using the convolution process. Multiple convolution kernels or filters are often applied and may be followed by the pooling operation, which helps to ensure that semantically similar features are merged to a single feature. Although this helps down-sampling, it also results in making CNNs spatially invariant. Hinton et al. (Hinton et al., 2011) observed that multiple layers of such down-sampling can lead to loss of spatial information and suggested the idea of "capsules" to characterize an entity, where neurons comprising a capsule capture both the spatial information and existence probability of the features into a vector.

Hinton's concept (Hinton et al., 2011) was practically applied by Sabour et al. (Sabour et al., 2017) by modifying CapsNets and proposing the dynamic routing by agreement algorithm to define the information flow from the lower to higher level capsules. This replaces the pooling operation in CNN. In CapsNets, capsules produce vectors rather than the scalar values in traditional CNN. The capsules in lower layers attempt to predict the output of each capsule in the next layer. The more these predictions agree, the more the output will be routed to the capsule in a higher layer.

This dynamic routing technique was found to be more efficient compared to pooling techniques, e.g., maxpooling, where only the most prominent outputs are routed to next layer. One or more convolutional layers feed the features into a capsule network. Consider a capsule network with two layers of capsules in layers $l$ and $l+1$, denoted $\boldsymbol{u}$ and $\boldsymbol{v}$, respectively. Initially, the output vector $u_i$ from the $i^{th}$ capsule in the lower level is sent to all higher-level capsules after scaling with the corresponding weight matrix $\boldsymbol{W}_{ij}$ to form the respective prediction vectors $\hat{u}_{j/i}$ as follows.

$$\hat{u}_{j/i} = W_{ij}.u_i \tag{1}$$

The total input $s_j$ to the $j^{th}$ higher level capsule is the sum of these predictions from the lower level multiplied by the corresponding coupling coefficients $c_{ij}$, which is expressed as follows.

$$s_j = \sum_i c_{ij}\, \hat{u}_{j/i} \tag{2}$$

The coupling coefficient between a lower-level capsule and all higher-level capsules must add up to 1 and is determined iteratively using the SoftMax operation on $b_{ij}$ as follows:

$$c_{ij} = \frac{\exp(b_{ij})}{\sum_k \exp(b_{ik})} \tag{3}$$

Where $b_{ij}$ represents the log prior probability responsible for determining if $u_i$ should be directed to $v_j$. The coupling coefficient increases if the predictions of the lower level agree with the output of the higher-level capsules. A non-linear squashing function (Eq. (4)) helps guarantee that the capsule's output lies between 0 and 1:

$$v_j = \frac{\parallel s_j \parallel^2}{1 + \parallel s_j \parallel^2} \frac{s_j}{\parallel s_j \parallel} \tag{4}$$

At the start of the iteration, the output from the lower capsule is directed to all higher capsules equally, and then $s_j$, $v_j$, and $c_{ij}$ are revised Using Eqs. (1)-(4). Then, $b_{ij}$ is updated as follows:

$$b_{ij} = b_{ij} + \hat{u}_{j/i}.v_j \tag{5}$$

where (.) represents a dot product. The coupling coefficients are updated based on the extent of agreement between $v_j$ and $\hat{u}_{j/i}$. This ensures that the higher level capsules receive more data from the lower-level capsules that they agree more with, thereby forming a part-whole association, which constitutes the main concept of the dynamic routing algorithm. Briefly, the capability of CapsNets to preserve the spatial context of the input features, can provide better discriminative ability than CNNs.

# 3. DATASETS

We focus on evaluating the capability of CapsNets in recognizing speakers from emotional speech and comparing the results to those of baseline models on three different speech datasets. These three datasets are summarized as follows.

### 3.1. Emirati Speech Database

The Emirati Speech Database (Shahin et al., 2018, 2019) was recorded at the College of Communication, University of Sharjah, United Arab Emirates and comprises of speech samples collected from 50 untrained speakers (25 male and 25 female; 14 to 55 years). This Arabic database includes eight commonly used sentences within the United Arab Emirates community expressed by the speakers both neutrally (devoid of emotion) and emotionally (involving five emotions: happy, sad, angry, fear, and disgust) with 9 repetitions for each instance. The time duration of audio samples spans 1–3 s. The recordings were acquired using a 16-bit linear A/D converter with a sampling rate of 44.6 kHz and later down-sampled to 12 kHz.

### 3.2. Ryerson Audio-Visual Database of Emotional Speech and Song (open-access)

RAVDESS (Livingstone & Russo, 2018), a multimodal dataset comprising recordings collected from 24 professionals (12 male and 12 female) has 7,356 files with vocalizations involving two sentences in North American accents. The speech data involve recordings where the sentences are expressed by the speakers with the following emotions: neutral, calm, happy, sad, angry, fearful, surprise, and disgust. Similarly, the calm, happy, sad, angry, and fearful emotions outline the song data. All emotions are recorded twice and grouped as normal and high based on their intensities, and all recordings are available in three forms, i.e., audio-only (16bits, 48kHz), audio-visual, and video-only (i.e., no sound). The audio-only files include 1,440 speech files and 1,012 song files. Note that audio-only recordings involving six emotions (i.e., neutral, happy, sad, angry, fear, and disgust) were used in this study.

### 3.3. Speech Under Simulated and Actual Stress Dataset

The SUSAS (J. H. Hansen et al., 1997) database, which was created by the Linguistic Data Consortium (LDC), comprises recordings with a broad range of emotions and stress. It contains over 16,000 audio files recorded under both actual and simulated stress collected from 32 speakers (19 male and 13 female; 22 to 76 years). The recordings were acquired using a 16-bit A/D converter with a sampling rate of 8 kHz. This study is based on a subset of the speech corpus involving 35 words gathered from 9 speakers spoken under the following conditions: neutral, soft, loud, anger, fast, and slow. The time duration of these audio samples spans 1–2 s.

## 4. CAPSNET-BASED EMOTONAL SPEAKER IDENTIFICATION MODEL

We investigated the performance of the proposed CapsNet architecture using MFCCs. The overview of the proposed emotional speaker identification model is shown in Fig. 1. The subsequent subsections briefly explore each of the blocks in the figure.

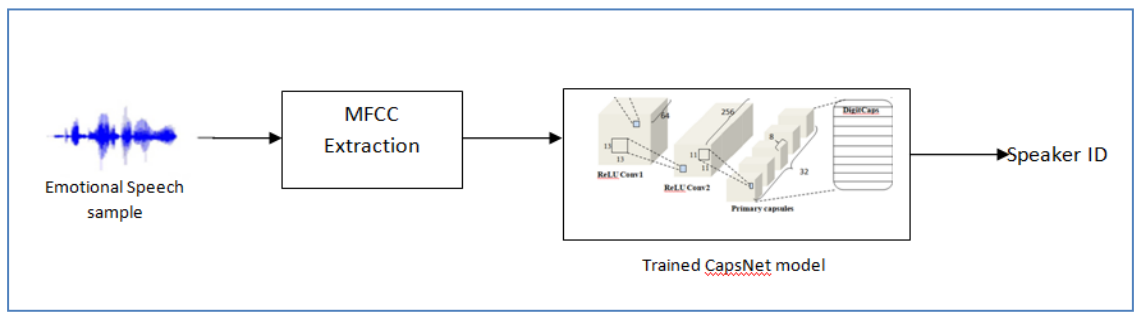

Fig. 1        Overview of the proposed emotional speaker identification model

### 4.1. Feature Extraction

MFCCs have been the go-to feature since their introduction in the 1980s and have been utilized extensively in many speech and speaker recognition studies (Tirumala et al., 2017). Computing MFCCs is a straightforward process, as shown in Fig. 2. This process begins by slicing audio samples into sliding frames of width 25 ms, resulting in statistically stationary signals. This is followed by discrete Fourier analysis of the frames to extract the frequency information. Then, the DFT power spectrum is computed and later scaled using the triangular mel-scale filter banks to a form mel-scale power spectrum, where each slot output represents the energy of the enveloped frequency bands, which is performed in an attempt to imitate the manner in which the human ear recognizes sounds.

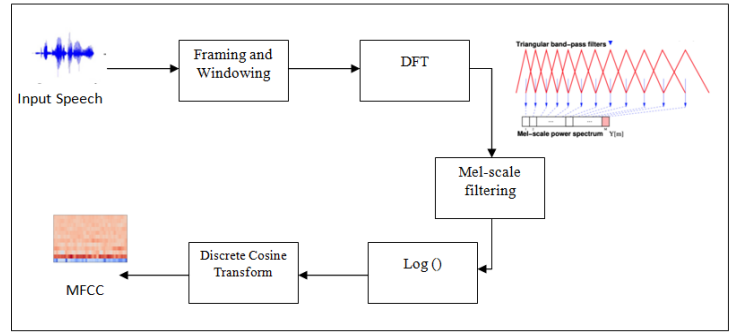

Fig. 2        Block diagram of MFCC computation

Human resistance to small variations at high energy compared to small changes at a low energy level can be considered as logarithmic. Thus, the log of the filter bank output is determined, followed by discrete cosine transformation to obtain MFCCs. Forty coefficients (i.e., 20 MFCCs and 20 delta MFCCs) were extracted per frame and stacked to form a two-dimensional matrix. In addition, zero padding was employed to overcome variation in the number of frames for different audio samples.

### 4.2. CapsNet Architecture

The CapsNet model proposed by Sabour et al. (Sabour et al., 2017) comprises two parts. The first part includes two convolutional layers with the final layer restructured to form the primary capsule layer followed by the digit capsule layer with capsules of 16 dimensions. The second part comprises a decoder network made of three fully-connected (FC) layers, which attempts reconstruction as a regularization technique. These two parts are merged to form the model, which is trained using a loss function comprising both margin loss ($L_c$), as given by Eq. (6), and reconstruction loss, which is a scaled version of the mean squared error (computed between the reconstructed output and input):

$$L_c = T_c \, max \, (0, m^+ - \parallel v_c \parallel)^2 + \lambda (1 - T_c) \, max(0, \parallel v_c \parallel - m^-)^2 \qquad (6)$$

where $v_c$ is the capsule output for class c. $T_c = 1$ if the target class is c,$m^+ = 0.9$, $m^- = 0.1$ and$\lambda = 0.5$. Thus, total loss (T) is given as follows.

$$T = \text{Margin loss} + \propto . \text{Reconstruction loss} \qquad (7)$$

$\propto = 0.0005$. The proposed model differs from the architecture described above with modifications introduced after considerable deliberations. After experimenting with different hyperparameters (including kernel size, convolution layers, and number of FC layers), the modified CapsNet architecture shown Fig. 3 was found to yield the best performance.

The structure of the modified proposed CapsNet architecture, hereby referred to as CapsNet-M (shown in Fig. 3) is described as follows. The first convolutional layer slides across the two-dimensional MFCC matrix using 64 kernels of size 15 with a stride of 1x5 and no padding, followed by the ReLU activation function. The second convolutional layer uses 256 kernels of size 13 with a stride of 1, which is also followed by the ReLU activation function. Next is the primary capsule layer, which is primarily a convolutional layer that uses a kernel size of 11 with a stride of 2 to form 256 feature maps of scalar values, which is then reshaped to form 32 feature maps comprising eight-dimensional vectors. Each of the 8D vectors are fully connected to the next layer comprising capsules with 16 dimensions to form the digit capsule (DigitCaps) layer with the number of outputs corresponding to the number of speakers. The dynamic routing algorithm helps to modify the coupling coefficients between the primary and digit capsules. Note that the length of the output vector provides the class probability. The decoder network of the proposed model is implemented using two FC layers and is fed by the digit capsule layer. However, training the CapsNet model involves masking the digit capsule outputs, except for the target class and for testing the masking is done based on the predicted labels. To learn better discriminative features, the model was trained using a loss function that combined margin loss and reconstruction error, as given by Eq. (7). This is in contrast to the cross-entropy error commonly used in CNNs. Note that margin loss also helps avoid overfitting.

We also designed three basic CapsNets that follow that proposed by Sabour et al. (Sabour et al., 2017) with one convolution layer followed by primary capsule and digit capsule layers. The first model is explained in the literature (Sabour et al., 2017) and uses 256 kernels of size 9 for the first convolutional layer. The second layer, which uses 256 kernels of size $9 \times 9$ (with a stride of 2), is restructured to form the primary capsule layer, which is then fully connected to the digit capsule layer. This model is referred to as the Caps-9 model.

The second model is similar to Caps-9, except for the first convolutional layer, which employs 256 kernels of size 15. The second layer is exactly the same as the that in Caps-9. This model is referred to as the Caps-15model. The third model, which is referred to as the Caps-19 model, is also similar to the Caps-9 model, except for the first convolutional layer, which employs 256 kernels of size 19. These models give us insight into the performance of the proposed CapsNet architecture with different hyperparameters.

### 4.3. Baseline Systems

An evaluation of the proposed CapsNet architecture was performed to examine whether the model can provide better results in identifying speakers using expressive or emotional audio samples by comparing other state-of-the-art classification models on the same datasets. A random forest classifier with 1,000 trees and a support vector machine were designed.

We also designed a baseline CNN model with four convolutional layers with ReLU nonlinearities and fully-connected layers on top. The first layer is a convolution layer that uses 128 kernels of size 13, followed by a maxpooling layer and a dropout layer. The second convolution layer uses 256 kernels of size 11, followed by a batch normalization layer and a maxpooling layer of size 2. The third layer is similar as before but has 256 kernels of size 5 and a stride of 2, followed by a maxpooling layer. The fourth layer uses 128 kernels of size 3. Next, is a global average pooling layer (Lin et al., 2013) that computes the mean value of each feature map and feeds to a SoftMax layer for final classification.

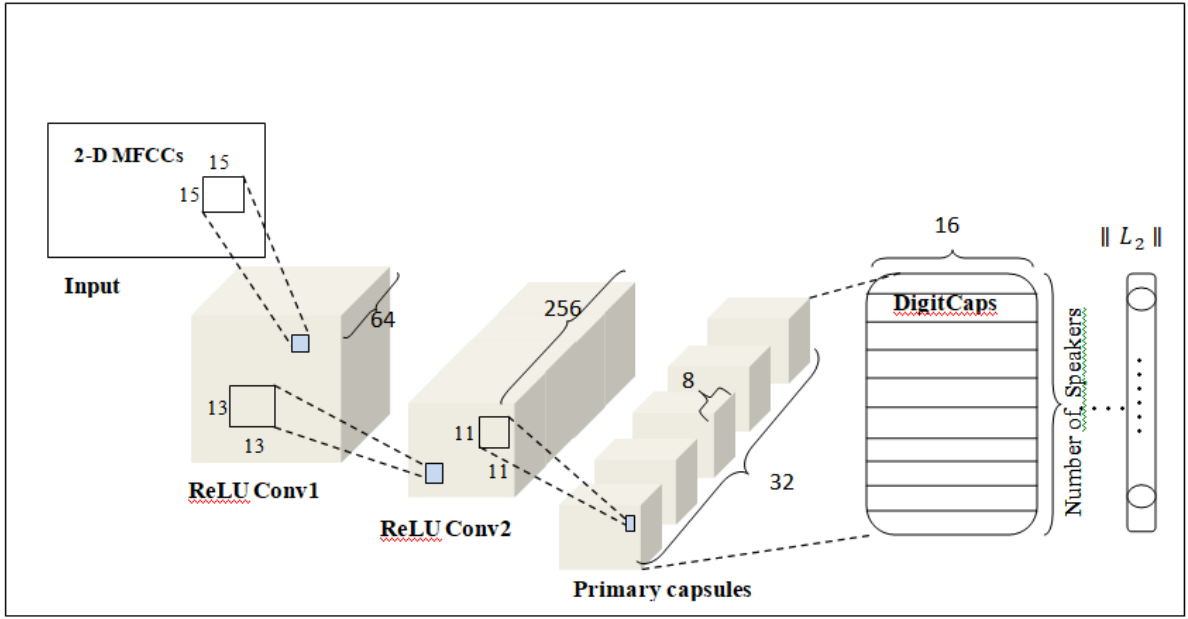

Fig. 3        Overview of proposed CapsNet-M architecture for text-independent speaker identification under emotional environment

### 4.4. Data Preparation and Experimental Setup

The models were evaluated using the Emirati Speech Database (ESD), RAVDESS, and SUSAS datasets. The audio files comprising these speech datasets were separated as explained below to train and test the models.

ESD comprises eight unique utterances (in different emotions, including neutral), of which neutral speech samples involving four utterances were selected to train the models, and the remaining four utterances were used to test the models. This was done to ensure that the experiments were performed in a text-independent context. Each trial was run five times during which the chosen utterances Were changed. During one trial, (50 speakers ×4 randomly chosen utterances × 1 neutral emotion × 9 repetitions) speech samples were used in the training phase and (50 speakers × 4 remaining utterances × 6 emotions × 9 repetitions) speech samples were used in the testing phase. The results were tabulated, and the trials were repeated except the utterances were changed. Note that neutral speech samples from four new randomly selected utterances were used in the next trial. The trials were repeated five times. In each scenario, the neutral speech samples from different combinations of utterances were selected to train the models. The results obtained during each trial were tabulated and averaged to obtain the final result, which was then analyzed.

RAVDESS comprises two unique utterances. Only samples corresponding to the neutral emotion of one utterance were used to train the models. The models were then evaluated in terms of identifying speakers using emotional speech (and neutral) samples using the second utterance, which ensured that the tests were conducted from a text-independent perspective.

The SUSAS dataset contains 35 unique words spoken in varying emotions, of which only samples corresponding to neutral emotion involving 15 words were used to train the models. The models were then evaluated based on their performance in identifying the speakers using emotional speech (and neutral) samples involving the remaining 20 words. The experiments were repeated five times, and the results obtained during each experiment were tabulated and averaged to obtain the final result.

The models were trained using the MFCCs extracted from the audio files, which were separated as explained previously. We used Keras with a TensorFlow (Abadi et al., 2016) backend for implementation, and training was performed using the Adam optimizer (Kingma & Ba, 2015) to minimize the loss given by Eq. (6). In addition, five-fold cross validation was employed in the experiments. A batch size of 64 and a learning rate of 0.001 were used for the hyperparameter settings.

## 5. EXPERIMENTAL ANALYSIS AND RESULTS

An evaluation of the proposed CapsNet-M model in classifying speakers using audio samples colored with emotions was conducted in several experiments, and the results were compared to the baseline systems. In all experiments, the models were trained using only neutral speech samples.

*Experiment 1*: The speaker identification performance of the proposed CapsNet-M model was compared to the random forest, SVM, and CNN baseline systems and Caps-9, Caps-15, and Caps-19 on the ESD. The results were also compared those obtained in a previous works using the GMM-DNN classifier (Shahin et al., 2018) on ESD. Accuracy is expressed as follows.

$$Accuracy\ (Acc) = \frac{Number\ of\ instances\ with\ correctly\ identified\ speaker}{Total\ number\ of\ trials} \times 100\% \quad (8)$$

The overall accuracy of the different models using audio samples with emotional content (anger, sad, fear, happy, disgust, and neutral) is summarized in Table1. As can be seen, all CapsNet architectures performed comparatively better than the baseline systems, with the proposed CapsNet-M model achieving 89.85% (some runs obtained overall performance as high as 91%).

CapsNet models are able to perform better than baseline systems due to their capability to exploit spatial association between the low-level features. CNNs, on the other hand, do not preserve the spatial context of the input features which results in information loss and thereby a lower accuracy. This can be verified by the results in Table 1 where the performance of CNN and GMM-DNN models was surpassed by all the four CapsNet models. Our results are conclusive with prior studies (*End-to-End Speech Command Recognition with Capsule Network*, 2018) using CapsNet models on speech recognition. A detailed evaluation of the performance of these models for different emotional circumstances is shown in Fig. 4.

Table 1.      Overall Accuracy of different classifiers on ESD

| Classifier | Random Forest | SVM | CNN | GMM-DNN | CAPS-9 | CAPS-15 | CAPS-19 | CAPSNET-M (Proposed) |
|---|---|---|---|---|---|---|---|---|
| Performance (%) | 72.52 | 75 | 78.52 | 81.7 | 84.34 | 87.01 | 87.30 | **89.85** |

As shown in Fig. 4, only the proposed CapsNet-M model obtained greater than 90% accuracy in identifying speakers with four different emotions, i.e., fear, disgust, sad, and neutral. Most classifiers are effective in identifying speakers when they communicate using neutral speech; however, only three models- the proposed CapsNet-M, Caps-15, and Caps-19 models obtained greater than 95%. In addition, the comparatively low scores obtained by all models when recognizing anger is attributed to variations in pitch, loudness, and other characteristics. Only the proposed model obtained greater than 75%identification rate in "Angry environment," while the CNN displays just over 60 percent in its performance Further detailed evaluation on the CapsNet-M model was performed using the precision, recall, F1 score, and area under the ROC curve (AUC) metrics. Precision, as given by Eq. (9), computes correct positive predictions, and recall gives an indication of missed positive predictions (Eq. (10)) by calculating the number of obtained correct positive predictions among the total positive predictions. Precision and recall can be combined to summarize the model performance using F1 score, as given by Eq. (11).

The precision, recall, and F1 score results obtained by the proposed CapsNet-M model were compared to those obtained using the CNN and SVM based on a subset of the ESD (Table 2). The proposed CapsNet-M model obtained an F1 score of ≥ 0.90 for nearly 75% of the speakers, while the CNN obtained an F1 score of ≥ 0.90 for less than 25% of the speakers.

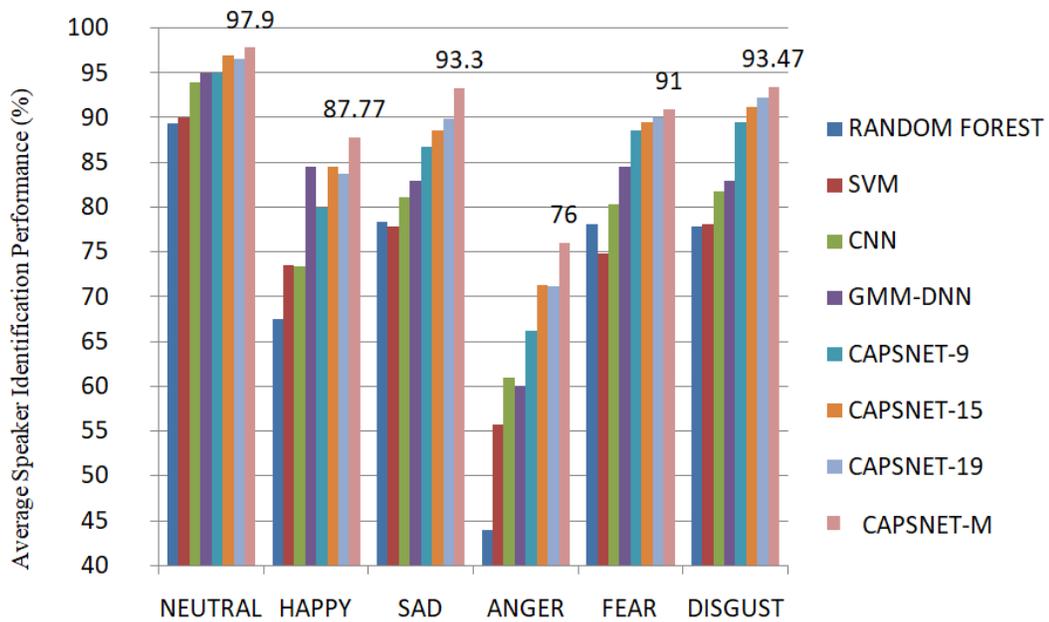

Fig. 4　　Performance under different emotional states based oneight classifiers on ESD

$$Precision = \frac{True\ Positives}{True\ Positives\ +\ False\ Positives} \qquad (9)$$

$$Recall = \frac{True\ Positives}{True\ Positives\ +\ False\ Negatives} \qquad (10)$$

$$F1 = 2 * \frac{Precision * Recall}{Precision + Recall} \qquad (11)$$

The AUC helps measure the area under the receiver operating characteristic curve plotted between the true positive and false positive rates at different thresholds. The closer the AUC score is to 1, the better the classifier. The AUC scores obtained by the proposed CapsNet-M, CNN, and SVM models were 0.993, 0.976, and 0.871, respectively, as shown in Table 3. The higher AUC score of the proposed CapsNet-M model indicates that it is a more "skillful" model compared to the other classifiers.

Table 2.        Precision, recall, and F1 score of proposed CapsNet, CNN, and SVM models

| Metrics | Observed number of speakers (of a total of 30 speakers) for different classifiers that satisfy the given metrics | | |
|---|---|---|---|
| | Proposed CapsNet-M | CNN | SVM |
| Precision≥ 0.90 | 22 | 12 | 11 |
| Recall≥ 0.90 | 22 | 8 | 8 |
| F1 Score ≥ 0.90 | 22 | 7 | 8 |

Table 3.        AUC scores of proposed CapsNet, CNN, and SVM models

| Area under the ROC curve | Proposed CapsNet-M | CNN | SVM |
|---|---|---|---|
| | **0.993** | 0.976 | 0.871 |

*Experiment 2*: The SUSAS database was used to evaluate the overall average accuracy of different models using audio samples involving six emotions, i.e., angry, loud, fat, slow, soft, and neutral, after training using neutral speech samples. The proposed CapsNet-M model was compared to other classification models, and the overall accuracy is summarized in Table 4. The performance of different classifiers and compensation methods with MFCCs as features was studied by Raja and Dandapat (Senthil Raja & Dandapat, 2010), and they recorded 53.6%, 53.96%, and 57.14% with the GMM,VQ, and selection of compensation by stress recognition (SCSR) techniques, respectively using the SUSAS dataset (involving angry, Lombard, neutral, and question).

Table 4.        Performance of different classifiers on SUSAS database

| Classifier | GMM(Senthil Raja & Dandapat, 2010) | SCSR(Senthil Raja & Dandapat, 2010) | SVM | CNN | CAPS-9 | CAPS-15 | CAPSNET-M (Proposed) |
|---|---|---|---|---|---|---|---|
| Accuracy (%) | 53.6 | 57.14 | 62 | 73.86 | 74.4 | 79.9 | **81.95** |

A detailed evaluation of the performance of the SVM, CNN, Caps-9, Caps-15, Caps-19 and proposed CapsNet-M models on the SUSAS dataset for different emotions (anger, loud, soft, fast, slow, and neutral) is shown in Table 5. Although the results demonstrate that the proposed CapsNet-M outperformed the compared models, it is clear that they struggled with identifying speakers with angry and loud speech samples.

Table 5.        Speaker identification accuracy (%) results of different emotional states using SUSAS dataset

| Classifier | Emotions | | | | | | |
|---|---|---|---|---|---|---|---|
| | Neutral | Angry | Loud | Soft | Slow | Fast | Average |
| SVM | 87.2 | 30.2 | 24.1 | 62.4 | 84.3 | 83.8 | 62.0 |
| **CNN** | 97.7 | 49.2 | 49.6 | 72.9 | 83.1 | **91.4** | 73.9 |
| **CAPS-9** | 97.0 | 47.14 | 47.1 | 81.2 | 85.1 | 88.3 | 74.4 |
| **CAPS-15** | **98.9** | 58.3 | 60.2 | **83.2** | **90.0** | 88.8 | 79.9 |
| **CAPS-19** | 97.2 | 55.5 | 58.6 | 81.3 | 85.9 | 89.3 | 77.8 |
| **CAPSNET-M (proposed)** | **98.9** | **63.2** | **65.0** | **83.2** | **90.0** | **91.4** | **82.0** |

*Experiment 3*: The performance of the proposed model was evaluated on the RAVDESS dataset, where the models were trained using only neutral speech from a single statement in both song and speech form. Table 6 compares the results obtained by the proposed CapsNet-M architecture to those of the random forest, SVM, CNN, and Caps-15 models in. The results obtained on RAVDESS are slightly lower (as expected) compared to the results obtained on other datasets due to the limited number of training samples. These results also help to evaluate the performance of the classifiers with a limited amount of training data. The results shown in Table 6 demonstrate that nearly all classifiers could identify the speakers under the neutral condition. However, the results wavered when tested on emotional speech, especially when the models were trained using a limited number of neutral speech samples.

Table 6.        Speaker identification (%) results of different emotional states on RAVDESS

| Classifier | Emotions | | | | | | |
|---|---|---|---|---|---|---|---|
| | Neutral | Happy | Sad | Anger | Fear | Disgust | Average |
| Random Forest | 92.5 | 57.9 | 61.7 | 34.2 | 45.2 | 38.5 | 55.0 |
| SVM | **99.0** | 58.5 | 60.1 | 45.7 | 49.4 | 47.9 | 60.1 |
| CNN | **99.0** | 76.0 | 74.4 | 50.0 | 59.5 | 51.0 | 68.3 |
| CAPS-15 | **99.0** | 75.5 | 76.5 | 61.7 | 66.4 | 54.1 | 72.2 |
| CAPSNET-M (proposed) | **99.0** | **77.6** | **78.7** | **65.0** | **67.0** | **62.5** | **75.0** |

*Experiment 4:* As the capsule network architecture is still in development, we feel it is vital to understand the effects of the number of routings and the reconstruction stage (with and without the decoder network) on the performance of these speaker identification models. The ESD was used, and 10 different experiments were conducted by varying the routings from 1 through 5, each with and without the reconstruction stage. The results shown in Fig. 5 indicate that the decoder network improved the performance of the model regardless of the number of routings. The results also indicate that applying two routing iterations worked sufficiently; however, optimal results were obtained using three routing iterations.

*Experiment 5:* The robustness of the proposed CapsNet-M classifier was evaluated in a noisy environment. Audio files of 30 speakers from the ESD were distorted with noisy signals at a ratio of 2:1 were used. The results in Table 7 demonstrate that although the performance of the CapsNet-M model was affected by the interference signals, it still outperformed the CNN in normal circumstances. In addition, the accuracy obtained on audio recordings in angry talking conditions was most affected, with the rates decreasing from 76% to 65% for the CapsNet model and 61% to 41% for the CNN model. To provide further insight, confusion matrix plots of the performances of the CapsNet-M model and CNN model in clean and noisy environment is shown in Fig. 6. The plots also show that the proposed

CapsNet model can better discriminate between speakers that the CNN model under both clean and noisy environments. The CapsNet has lighter shades along the diagonal which indicates better performance. While the CNN has darker shades and more misclassification as shown in the plots.

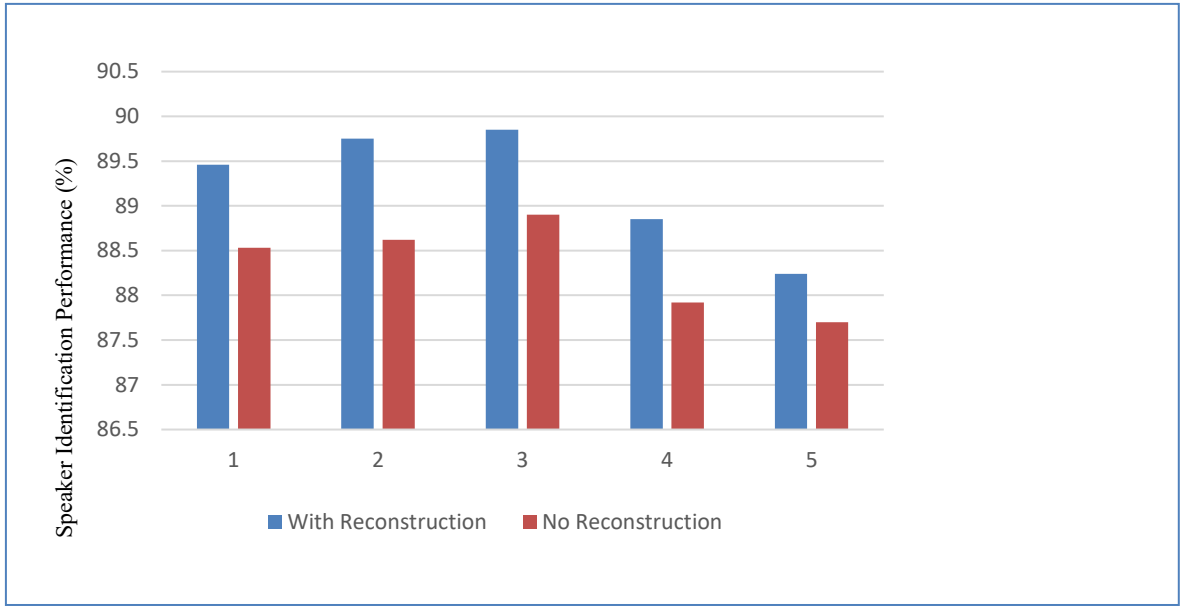

Fig. 5        Effects of varying number of routing iterations and reconstruction phase on accuracy on ESD

*Experiment 6:* The training and testing times of the CNN and CapsNet-M models were compared. For the CNN, a single epoch required approximately 7–12 s, and training with approximately 250–300 epochs (Fig. 8) required 30 to 50 minutes. In contrast, a single epoch required 10–20 s for the CapsNet-M network (Fig. 7). However, due to the fast convergence of the proposed CapsNet-M model (Fig. 7), training only required 30 (or a maximum of 40) epochs, which took a maximum of 12 minutes. Note that tests were run on Google Colab using Tesla K80 and the learning rate was chosen as 0.001. This is demonstrated in Table 8.

Table 7.        Performance (%) on distorted speech with proposed CapsNet-M and CNN models

| Emotion | CapsNet-M | | CNN | |
|---|---|---|---|---|
| | Normal | Distorted | Normal | Distorted |
| Neutral | 97.4 | 92.0 | 92.4 | 82.7 |
| Happy | 87.8 | 78.3 | 67.2 | 62.1 |

| Sad | 92.0 | 88.0 | 84.7 | 71.2 |
| Anger | 76.0 | 65.0 | 52.4 | 43.1 |
| Fear | 90.5 | 78.9 | 74.7 | 63.0 |
| Disgust | 93.4 | 88.6 | 86.2 | 74.7 |
| Average | 89.5 | 81.8 | 76.2 | 66.1 |

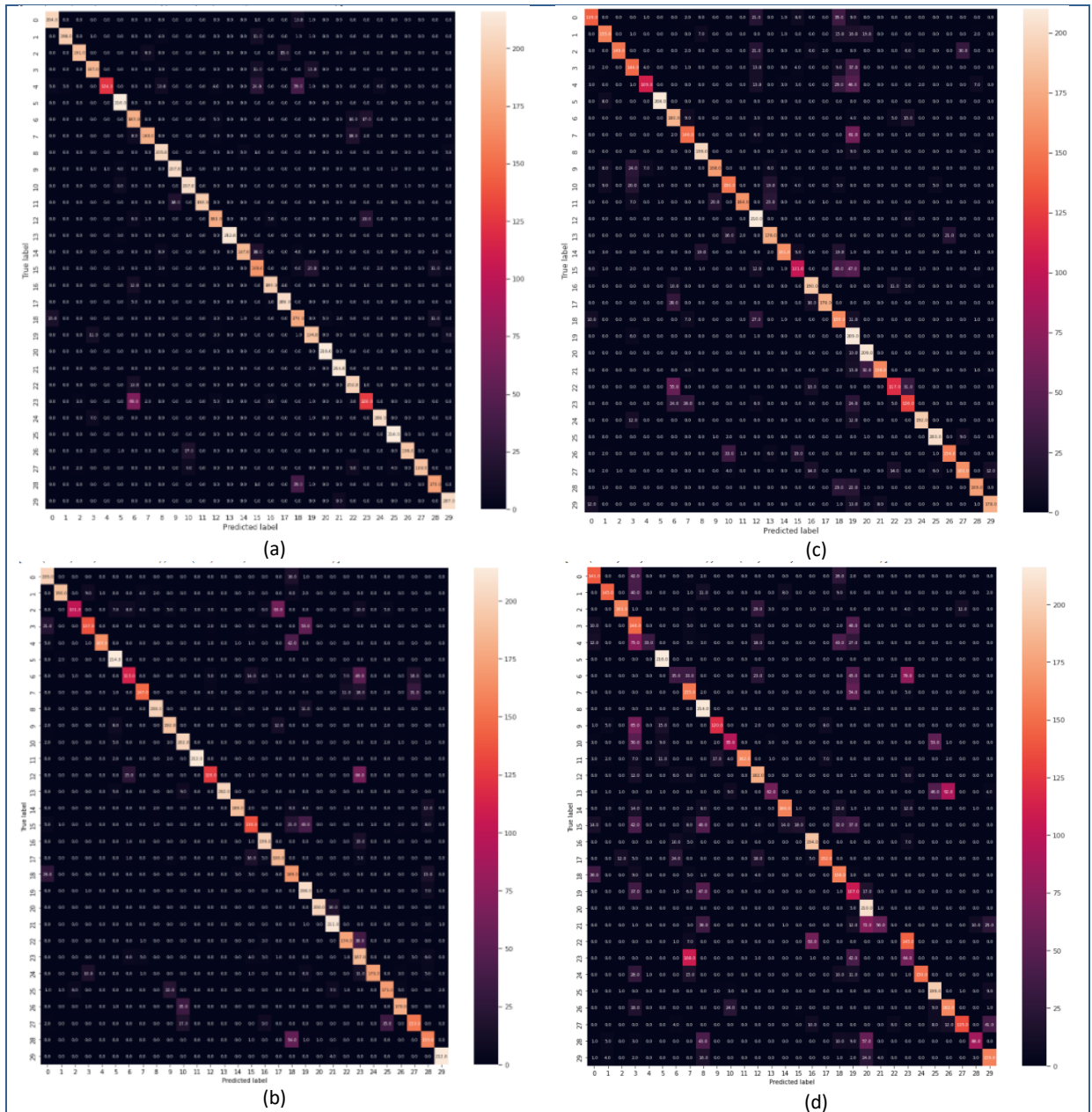

Fig. 6    Confusion matrix plots obtained for CapsNet-M and CNN models under clean and noisy environments using Emirati Speech Dataset. (a) CapsNet-M in clean environment, (b) CapsNet-M in noisy environment, (c) CNN in clean environment and (d) CNN in noisy environment.

Table 8.        Comparison of training time for CapsNet-M and CNN models

| Models | CapsNet-M | CNN |
|---|---|---|
| Time for single epoch | 10-20 sec | 7-12 sec |
| Total no: of epochs needed | 30-40 | 250-300 |
| Total training time | 12 min | 30 – 50 min |

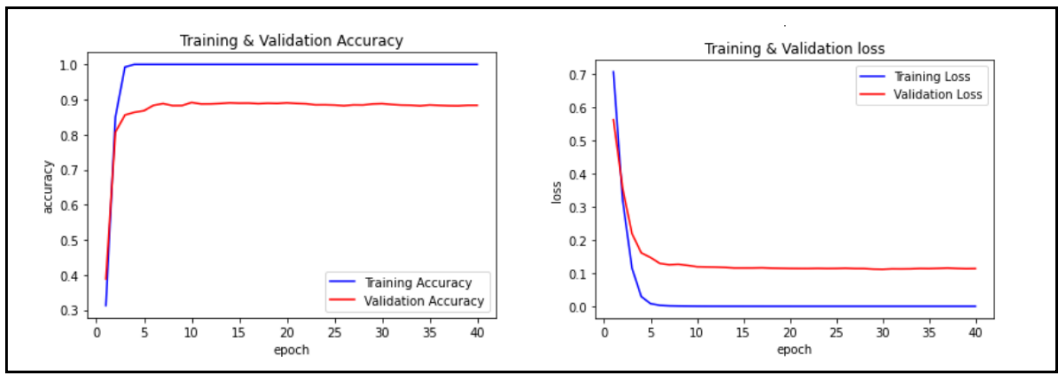

Fig. 7        Accuracy and loss variations of proposed CapsNet-M model during training and validation over 40 epochs

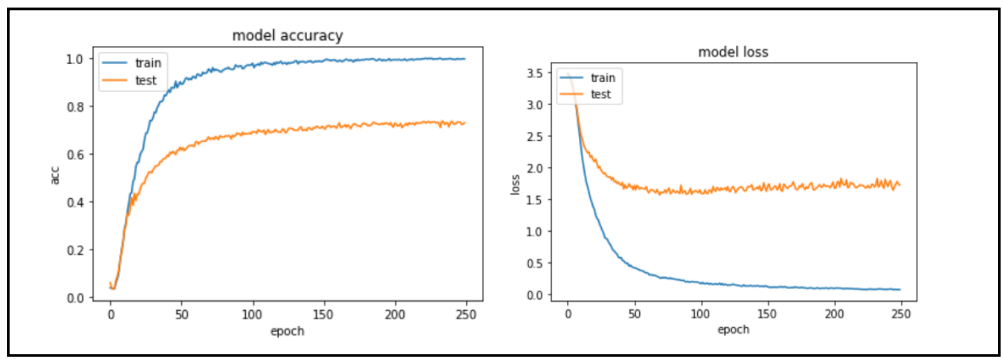

Fig. 8        Accuracy and loss variations of CNN during training and testing

*Experiment 7:* Further evaluation of the proposed model was performed using statistical significance analysis, which is frequently performed to ensure there is a significant distinction between the models by exploiting the p-value test measure. The p-value represents the probability under which the statistical result occurs and should be less than the significance level for the results to be considered statistically significant or, as in our case, the models to be statistically distinct, with the customary confidence level being 95% (i.e., a significance level of 5%).

The non-parametric Wilcoxon signed-rank test was used to evaluate the different models. The results shown in Table 9 suggest that the proposed CapsNet-M architecture is statistically different from the other models because the p-values for each pair is less than 0.05.

Table 9.        P-value obtained using Wilcoxon signed-rank test for proposed CapsNet and other classifiers

| Classifier | CapsNet-M | CNN | SVM | Random Forest |
|---|---|---|---|---|
| **CapsNet-M** | ** | 0.043 | 0.043 | 0.042 |

As a comparison between our proposed model with previous work, we faced some challenges due to the limitations of emotional speaker identification related work that uses the same datasets. However, we list below a few previous related work that can be compared with our proposed model.

Our proposed model gives an accuracy of 89.85% (average of all emotions) using the Emirati Accented Database which outperforms the work in (Nassif et al., 2021) where the best accuracy obtained was 83.7% and the work in (Shahin et al., 2018) where the best accuracy obtained was 81.7%. On the other hand, our proposed model gives an accuracy of 99% in neutral condition based on RAVDESS dataset. This outperforms the work in (Sefara & Mokgonyane, 2020) which is 92% based on RAVDESS.

# 6. CONCLUSIONS AND LIMITATIONS

Speech samples procured under highly emotional circumstances degrade the performance of speaker identification models, and training models using speech samples with varying emotional content can improve performance but is not a viable solution. Thus, in this paper, we have proposed a novel text-independent speaker identification model based on the capsule network architecture. The proposed architecture was evaluated on the Arabic ESD, English SUSAS database, and English RAVDESS database, and the results demonstrate that the proposed architecture outperforms the random forest, SVM, MLP, GMM-DNN, and CNN models relative to accuracy, precision, recall, and AUC. In our experiments, the proposed CapsNet-M model was trained using neutral speech samples and evaluated on emotional speech samples with MFCCs as the only input feature. Overall, the proposed CapsNet-M architecture obtained higher average speaker identification rates compared to the baseline systems. In addition, the proposed model outperformed the CNN technique and required less training time.

This work has some limitations. For instance, average Speaker identification accuracy based on the proposed model is non ideal. The lowest accuracy takes place when speakers talk angrily. Although we used quite large databases, the number of speakers is still limited.

Moreover, the current research only uses MFCC as the input features. It would be beneficial to examine the accuracy with different input features such as spectrograms.

Thus, in future, we plan to improve performance for the angry emotional state. There is still room to improve speaker recognition from audio samples in angry emotional situations. In addition, we plan to study speaker identification performance in emotional environments using different input features, e.g., spectrograms. Finally, further study into employing capsule networks in speaker and speech recognition with proven results is required because capsule networks are currently in development.

**Compliance with Ethical Standards**

The authors thank the University of Sharjah for supporting this work through the Machine Learning and Arabic Language Processing research group. Dr Adi Alhudhaif is thankful to the Deanship of Scientific Research at Prince Sattam bin Abdulaziz University, Al-kharj, Saudi Arabia.

**Conflict of Interest:** The authors declare that they have no conflict of interest.

**Informed Consent:** This study does not involve any experiments on animals.